\documentclass[aps,pra,showpacs,amsmath,amssymb,amsfonts,twocolumn,lengthcheck,superscriptaddress]{revtex4-1}
\usepackage[colorlinks=true,citecolor=blue,linkcolor=blue,urlcolor=blue]{hyperref}

\usepackage{graphicx}
\usepackage{subfigure}
\usepackage{verbatim}
\usepackage{bm}
\usepackage{epsf}
\usepackage{color}
\usepackage{hhline}
\usepackage{mathtools}

\newcommand{\bra}[1]{\left\langle #1\right|}
\newcommand{\ket}[1]{\left|#1\right\rangle}

\newcommand{\tr}[1]{\mathrm{tr}\left\{#1\right\}}

\newcommand{\la}{\left\langle}
\newcommand{\ra}{\right\rangle}

\newcommand{\mi}[1]{\min{\left\{#1\right\}}}

\newcommand{\e}[1]{\exp{\left(#1\right)}}
\newcommand{\lo}[1]{\ln{\left(#1\right)}}

\newcommand{\co}[1]{\cos{\left(#1\right)}}

\newcommand{\bla}{bla\\bla\\bla\\bla\\bla}

\newcommand{\PRA}{Phys. Rev. A }

\newcommand{\mc}[1]{\mathcal{#1}}

\newcommand{\mrm}[1]{\mathrm{#1}}

\begin{document}

\title{Quantum speed limits and the maximal rate of quantum learning}

\author{Thiago V. Acconcia} 
\email[]{thiagova@ifi.unicamp.br} 
\affiliation{Instituto de F\'\i sica Gleb Wataghin, Universidade Estadual de Campinas - Unicamp,  Rua S\'ergio Buarque de Holanda, 777,  13083-859 Campinas, SP, Brazil}

\author{Sebastian Deffner} 
\email[]{deffner@umbc.edu}
\affiliation{Department of Physics, University of Maryland Baltimore County, Baltimore, MD 21250, USA}


\date{\today}

\begin{abstract} 
The Bremermann-Bekenstein bound is a fundamental bound on the maximal rate with which information can be transmitted. However, its derivation relies on rather weak estimates and plausibility arguments, which make the application of the bound impractical in the laboratory. In this paper, we revisit the bound and extend its scope by explicitly accounting for the back action of quantum measurements and refined expressions of the quantum speed limit. Our result can be interpreted as an upper bound on the \emph{maximal rate of quantum learning}, and we show that the Bremermann-Bekenstein bound follows as a particular limit. Our results are illustrated, by first deriving a tractable expression from time-dependent perturbation theory, and then evaluating the bound for two time-dependent systems -- the harmonic oscillator and the P\"oschl-Teller potential.
\end{abstract}

\maketitle

\section{Introduction\label{sec:Intro}}

Recent breakthroughs in nanotechnology have led to the development of smaller and more powerful devices \cite{Cotter1999,Rooks2006}, which mark the advent of the post-Moore era \cite{Colwell2013} of the \emph{Information age} \cite{Castells2011}. In particular, the last few years have seen the first industrial attempts at making (semi-)quantum computers publicly available, such as the DWave system \cite{D-Wave2016} and IBM's quantum experience \cite{IBM2016}. Generally, quantum computers are expected to provide  an exponential speed-up over classical architectures \cite{Feynman1982,Ronnow2014,Boixo2016} for certain tasks such as to factorize  large numbers \cite{sho99} or to search through unsorted databases \cite{Grover1996,Grover1997}.

However, as Landauer pointedly remarked ``information is physical'' \cite{landauer_1991}, and hence also quantum computers are subject to the fundamental laws of physics such as thermodynamics, special relativity, and quantum mechanics \cite{Bremermann1967}. In order to be useful in practical applications, it will be inevitable for quantum computers  to communicate with their classical environment. Hence, the natural question arises whether fundamental principles such as the uncertainty relations set constraints on the rate with which quantum information can be communicated. The Bremermann-Bekenstein bound \cite{Bremermann1967,Bekenstein1981,bek81,bek84} is an estimate for the upper bound on the rate of information transmission, which is defined as the ratio of the maximal amount of information stored in a given region of space divided by the quantum speed limit time \cite{Deffner2017Review}. The quantum speed limit is the maximal rate with which a quantum system can evolve, and it can be understood as a physically sound formulation of the uncertainty relation for energy and time \cite{Deffner2017Review}.

Although conceptually insightful, the Bremermann-Bekenstein bound can neither be considered satisfactory and nor practical for applications in quantum computing. Its derivation explicitly assumes that the complete information stored in a quantum system is accessible, i.e., it neglects the loss of information due to the back action of generic quantum measurements \cite{Nielsen2010}.

In this paper, we will revisit the Bremermann-Bekenstein bound and propose its generalization to include the effect of quantum measurements. To this end, we will study the \emph{maximal rate of quantum learning}, which is given by the ratio of the change of accessible information \cite{Holevo1973b} during a small perturbation divided by the quantum speed limit time \cite{Deffner2017Review}. We will see that the original Bremermann-Bekenstein bound is included in our approach as a special case. The general case, however, is mathematically rather involved, and thus we will express the maximal rate of quantum learning by  means of time-dependent perturbation theory. Our general results will then be illustrated for two experimentally relevant case studies, namely the driven harmonic oscillator and the P\"oschl-Teller potential.

\section{Notions and definitions}
\paragraph*{Bremermann-Bekenstein bound.}

The fundamental laws of physics govern the modes of operation of any computer \cite{Lloyd2000} and, thus, the processing of information \cite{Bremermann1967,Bremermann1982}. Bremermann suggested that information processes are limited by  three physical barriers: the light, the quantum, and the thermodynamic barrier. The light barrier is a consequence of  special relativity \cite{ein05_2}, which bounds the rate of transmission by the speed of light. The quantum barrier arises from Shannon's definition for the capacity of a continuous channel, $C_{max} = mc^{2}/h$ \cite{Bremermann1967}, which expresses that the maximum channel capacity is proportional to the mass of the computer. The latter can also be interpreted as a limit imposed by the first law of thermodynamics. Finally, the second law asserts that entropy of isolated systems cannot decrease. Thus, when $I$ bits of information are encoded, the probability of a given state decreases by $2^{-I}$, and consequently the entropy decreases by a factor $I \ k_{B} \ln 2$. 

However, it was quickly noted that this argumentation is somewhat dubious, since equating the maximal amount of information processed in a computation cannot  be fully described by Shannon's channel capacity. Therefore, Bekenstein revisited the issue from a cosmological point-of-view \cite{Bekenstein1981,bek81,bek84}. Starting from an upper bound on the information encoded in a system with energy $E$, Bekenstein derived the maximal rate with which information can be transmitted \cite{bek81},
\begin{equation}
\label{eq01}
\dot{I}\leq\dfrac{I}{\tau_\mrm{QSL}} \leq \dfrac{\pi E}{\hbar \ln2}\ ,
\end{equation}
where $E$ is the energy in the receiver's frame and $\tau_\mrm{QSL}$ is the minimal time necessary to transmit this information, i.e., the quantum speed limit time \cite{man45,Margolus98,Deffner2017Review}.

It is worth emphasizing that although insightful the Bremermann-Bekenstein bound \eqref{eq01} is a rather weak upper limit on the rate with which  information can be transmitted, or entropy be produced in a quantum system \cite{Deffner2010}. The reason is that in Eq.~\eqref{eq01} the total information stored in a quantum system is assumed to be accessible. This is generally not the case, since accessing information is accompanied by the back-action of quantum measurements \cite{Nielsen2010} -- in simple terms ``the collapse of the wave-function''.

\paragraph*{Accessible information.}

For the sake of simplicity, imagine that we have only access to a projective observable $A=\sum_\alpha a_\alpha \Pi_\alpha$, where $a_\alpha$ are the measurement outcomes, and $\Pi_\alpha$ are the projectors into the eigenspaces corresponding to $a_\alpha$. Typically, the post-measurement quantum state $\rho$ suffers from a back-action, i.e, information about the quantum system is lost in the measurement \cite{Groenewold1971}. How much information is lost is quantified by Holevo's information \cite{Holevo1973b},
\begin{equation}
\label{eq02}
\chi=S(\rho)-\sum_\alpha p_\alpha\, S(\rho_\alpha)\,,
\end{equation}
where $S(\rho)=-\tr{\rho \lo{\rho}}$ is the von-Neumann entropy, and $\rho_\alpha=\Pi_\alpha \rho \Pi_\alpha$ is the post-measurement state. Further, $p_\alpha=\tr{\Pi_\alpha \rho}$ denotes the probability to obtain the $\alpha$th measurement outcome. Note that the present arguments readily generalize to arbitrary POVM's instead of projective measurements \cite{Nielsen2010,kafri2012}.

The Holevo information \eqref{eq02} is always non-negative $\chi \geq 0$, which follows from the concavity of the entropy \cite{Nielsen2010}. However any perturbation of the system leads to a change in Holevo information, which is always nonpositive, $\Delta \chi \leq 0$, \cite{Preskill1998}. In other words, any perturbation allows to access more information than retrieved by the first measurement.

Therefore, $\chi$ \eqref{eq02} is our natural starting point to define ``quantum learning''. Any change in $\chi$ due to some controlled perturbation decreases our ignorance about the quantum system. We will further assume that all perturbations can be expressed as unitary maps $U$, and thus we can write for the change of $\chi$ during time $\tau$,
\begin{equation}
\label{eq03}
\Delta \chi = - \sum_\alpha p_\alpha(\tau)\, S(\rho_\alpha(\tau)) + p_\alpha(0)\, S(\rho_\alpha(0))\ .
\end{equation} 
Note that restricting ourselves to unitary perturbations is a judicious choice, since this guarantees that the total information, $S(\rho)$, remains unaffected by the perturbation. Hence, the only effect of the perturbation is a change in \emph{accessible} information. To now define the maximal rate of quantum learning we need the quantum speed limit time, $\tau_\mrm{QSL}$.

\paragraph*{Quantum speed limit.}

The quantum speed limit originally arose by a careful derivation of Heisenberg's uncertainty relation of energy and time \cite{man45,Margolus98}. More recently, it has found applications in virtually all areas of quantum physics, and the quantum speed limit has become an active area of research, see a recent review \cite{Deffner2017Review} and references therein. 

For driven systems, the quantum speed limit time from the geometric approach has proven to be practical \cite{Deffner2013PRL},
\begin{equation}
\label{eq04}
\tau_\mrm{QSL} = \dfrac{\hbar}{2 E_{\tau}} \sin^2[\mathcal{L}(\rho(0), \sigma(\tau))] \ ,
\end{equation}
where $E_{\tau} = \mi{1/\tau\, \int_0^{\tau}dt\, \| \rho(t) \mathcal{H}(t) \|_{p}} $, and $\mc{H}$ is the Hamiltonian generator of the unitary map $U=\e{-i/\hbar\,\int_0^\tau dt\,\mc{H}(t)}$. The norm is given by $\|A\|_p = (\tr{\vert A \vert^{p}})^{1/p}$, with $p \in \lbrace 1,2,\infty\rbrace$ \cite{Deffner2013PRL}. Further, $\mc{L}$ denotes the Bures angle,
\begin{equation}
\label{eq05}
\mc{L}(\rho(0), \sigma(\tau))=\arccos\left(\sqrt{\bra{\Psi_0}\sigma(\tau)\ket{\Psi_0}}\right)\,,
\end{equation}
where we assume for the sake of simplicity that the initial state is pure, $\rho(0)=\ket{\Psi_0}\bra{\Psi_0}$.

Generally, the operator norm, $p=\infty$, gives the sharpest bound \cite{Deffner2013PRL}, however the Hilbert-Schmidt or Frobenius norm, $p=2$, behaves qualitatively similiar, and it is significantly easier to compute \cite{Deffner2017Review,Deffner2017}. Thus for the sake of simplicity, we will work in the following with the quantum speed limit time expressed in terms of the Hilbert-Schmidt norm.

\paragraph*{Generalized Bremermann-Bekenstein bound.}

Having established all the ingredients, we can now move on to define the \emph{maximal rate of quantum learning}. To this end, imagine the following situation: an experimentalist obtains a quantum state $\rho_0$ and they have access to a quantum observable $A$, such as magnetization, parity etc. Generally the experimentalist can then retrieve $\chi$ \eqref{eq02} bits of information about the quantum state. Now lets further assume that the experimentalist can unitarily perturb the quantum system, and measure the same observable $A$ again. Due to this perturbation the additional amount of $\Delta \chi$ \eqref{eq03} has become accessible. The rate with which $\chi$ changes is upper bounded by the quantum speed limit,
\begin{equation}
\label{eq06}
\vert\dot{\chi}\vert \leqslant \dfrac{\vert \Delta \chi\vert}{\tau_\mrm{QSL}} \equiv |\Omega|\,,
\end{equation}
where we call $\Omega$ the \emph{maximal rate of quantum learning}. We stress again that the present arguments readily apply to POVMs, and we are not restricted to projective measurements.

It is easy to see that Eq.~\eqref{eq06} is a generalization of the Bremermann-Bekenstein bound accounting for accessibility of quantum information. To this end, assume that the post-measurement state can be written in terms of some inverse temperature $\beta$ as $\rho_\alpha=\e{-\beta H_\alpha}$. Then we can write Eq.~\eqref{eq03} as
\begin{equation}
\label{eq07}
\Delta \chi=\beta\left[\sum_\alpha \left(p_\alpha(\tau) \la H_\alpha(\tau)\ra-p_\alpha(0) \la H_\alpha(0)\ra\right)\right]\,.
\end{equation}
Thus we further obtain,
\begin{equation}
\label{eq08}
|\dot{\chi}|\leq|\Omega|=\frac{\beta \Delta E}{\tau_\mrm{QSL}}\,,
\end{equation}
where $\Delta E=\la H(\tau)\ra-\la H(0)\ra$, and $H_\alpha =\Pi_\alpha H \Pi_\alpha$. Equation~\eqref{eq08} is formally identical to the generalized Bremermann-Bekenstein bound derived in Ref.~\cite{Deffner2010} from quantum thermodynamic considerations.

\section{Illustrative case studies}

The remainder of this analysis is dedicated to illustrative case studies of the maximal rate of quantum learning \eqref{eq06}. For non-trivial systems computing both $\chi$ and $\tau_\mrm{QSL}$ quickly becomes mathematically involved. Therefore, we first derive an expression for $\Omega$ by means of time-dependent perturbation theory.

\subsection{Time-dependent perturbation theory}

Consider an arbitrary time-dependent Hamiltonian,
\begin{equation}
\label{eq09}
\mathcal{H}(t) = H_0 + \delta \lambda \ V(t) \ ,
\end{equation}	
where $H_0$ describes the unperturbed system, and $V(t)$ is an arbitrary time-dependent perturbation that plays the role of an external agent of small amplitude $\delta\lambda$. Then the unitary time-evolution operator can be written as a Dyson series \cite{cohen_vol2_77} in linear order of $\delta\lambda$ as,
\begin{equation}
\label{eq10}
\tilde{U}(t) \equiv \left(\mathbb{I} - i \dfrac{\delta\lambda}{\hbar} \int_0^{t} ds\,e^{\frac{i}{\hbar} H_0 s} V(s) e^{-\frac{i}{\hbar} H_0 s} \right)e^{-\frac{i}{\hbar} H_0 t} \,.
\end{equation}
Note $\tilde{U}(t)$ is the truncated Dyson series for the time-evolution operator in the Schr\"{o}dinger picture, and that we have $\delta\lambda \rightarrow 0: \tilde{U}(t) \rightarrow U(t) = e^{-\frac{i}{\hbar} H_0 t}$, recovering the unperturbed expression of the time-evolution operator. In the following, it will be convenient to adopt the notation: $\mathcal{I}(t) = \int_0^{t} ds\, e^{\frac{i}{\hbar} H_0 s} V(s) e^{-\frac{i}{\hbar} H_0 s}$.

Strictly speaking $\tilde{U}(t)$ is not a valid time-evolution operator, since due to the truncation the normalization of the time-evolved states is violated. Therefore, we need to introduce a normalization function $N(t)$,
\begin{equation}
\label{eq11}
\langle \Psi(t) \vert \Psi(t) \rangle = \dfrac{1}{N(t)} \langle \Psi_0 \vert \tilde{U}(t)^{\dagger} \tilde{U}(t) \vert \Psi_0 \rangle = 1\,,
\end{equation}
from which we obtain
\begin{equation}
\label{eq12}
\begin{split}
&N(t) = 1 + i \dfrac{\delta\lambda}{\hbar}\langle(\mathcal{I}^{\dagger}(t)-\mathcal{I}(t))\rangle_{\Phi}\\ 
&\quad+ \left(\dfrac{\delta\lambda}{\hbar}\right)^{2} \langle\mathcal{I}^{\dagger}(t) \mathcal{I}(t) \rangle_{\Phi}\ ,
\end{split}
\end{equation}
where $\langle \cdot \rangle_{\Phi} = \langle \Phi(t) \vert \cdot \vert \Phi(t) \rangle $, $\vert \Phi(t) \rangle = U(t) \vert \Psi_0 \rangle$. Accordingly, the time-dependent density operator becomes
\begin{equation}
\label{eq13}
\begin{split}
&\rho(t) = \vert \Psi(t) \rangle\langle \Psi(t) \vert  \\
&=\dfrac{1}{N(t)}\left(\mathbb{I} - i \dfrac{\delta\lambda}{\hbar} \mathcal{I}(t)\right)e^{-\frac{i}{\hbar} H_0 t}\rho_0 e^{\frac{i}{\hbar} H_0 t}\left(\mathbb{I} + i \dfrac{\delta\lambda}{\hbar} \mathcal{I}^{\dagger}(t)\right)\\
&= \rho^\mrm{in}(t) +  \delta\rho(t)\,,
\end{split}
\end{equation}
where $\rho^\mrm{in}(t) = e^{-\frac{i}{\hbar} H_0 t}\,\rho_0\,e^{\frac{i}{\hbar} H_0 t} / N(t)$, and $\delta\rho(t) = i \delta\lambda \ (\rho^\mrm{in}(t) \mathcal{I}^{\dagger}(t) - \mathcal{I}(t)\rho^\mrm{in}(t))/\hbar$, and where we only collected terms which are at most linear in $\delta\lambda$.
	
\paragraph*{Change of accessible information.}

From Eq.~\eqref{eq13} it is then straight forward to compute $\Delta\chi$ \eqref{eq03}. In particular, the probability to obtain measurement outcome $\alpha$ after the perturbation, $p_{\alpha}(\tau)$ reads
\begin{equation}
\label{eq14}
\begin{split}
&p_{\alpha}(\tau) = \tr{\Pi_{\alpha} \rho(\tau)} \\
&\quad= \sum_{k, \alpha} \langle k \vert \ \lbrace \vert \alpha\rangle\langle \alpha\vert \ (\rho^\mrm{in}(\tau) +  \delta\rho(\tau) )\rbrace \ \vert k \rangle\\ 
&\quad= p^\mrm{in}_\alpha(\tau) + \tr{ \delta\rho_\alpha(\tau)}\,.
\end{split}
\end{equation}
where $p^\mrm{in}_\alpha(\tau) = \tr{\Pi_\alpha \rho^\mrm{in}(\tau)}$. Similarly, the von-Neumann information of the post-measurement state $\rho_\alpha(\tau)$ becomes
\begin{equation}
\label{eq15}
\begin{split}
&S(\rho_\alpha(\tau)) = -\tr{\rho_\alpha(\tau) \log \rho_\alpha(\tau)}\\
& \quad= -\tr{ (\rho^\mrm{in}_\alpha(\tau) +  \delta\rho_\alpha(\tau)) \log (\rho_\alpha^\mrm{in}(\tau) +  \delta\rho_\alpha(\tau)) } \\
&\quad = S(\rho^\mrm{in}_\alpha(\tau))-\tr{\delta\rho_\alpha(\tau)} - \tr{\delta\rho_\alpha(\tau) \log  \rho^\mrm{in}_\alpha(\tau)}.
\end{split}
\end{equation}
Finally, collecting Eqs.~\eqref{eq14}-\eqref{eq15} and substituting into \eqref{eq03} an approximate expression for $\Delta\chi$ can be written as
\begin{equation}
\label{eq16}
\begin{split}
\Delta \chi^\mrm{lin} &= \sum_{\alpha} \Delta\chi^\mrm{in}_\alpha + \tr{\delta\rho_\alpha(\tau)} [p^\mrm{in}_\alpha(\tau) - S( \rho^\mrm{in}_\alpha(\tau)) ]\\
&\quad + p^\mrm{in}_\alpha(\tau) \tr{\delta\rho_\alpha(\tau) \log  \rho^\mrm{in}_\alpha(\tau)} ,
\end{split}
\end{equation} 
where $\Delta\chi^\mrm{in}_\alpha = -p^\mrm{in}_\alpha(\tau) S( \rho^\mrm{in}_\alpha(\tau)) + p_\alpha(0) S( \rho_\alpha(0))$. 
	
\paragraph*{Quantum speed limit.}

Having derived an expression for the numerator of $\Omega$ \eqref{eq06}, we continue by also expressing the quantum speed limit time, $\tau_\mrm{QSL}$ \eqref{eq04}, by means of perturbation theory. The time-dependent fidelity, $F=[\co{\mc{L}}]^2$, can be written as	
\begin{equation}
\label{eq17}
F(\rho(0), \rho(\tau)) \simeq \sqrt{\langle \rho^\mrm{in}(\tau)\rangle_{\Psi_0}} + \dfrac{1}{2} \dfrac{\langle\delta\rho(\tau)\rangle_{\Psi_0}}{ \sqrt{\langle \rho^\mrm{in}(\tau)\rangle_{\Psi_0}}}
\end{equation}
where $\langle \cdot \rangle_{\Psi_0} = \langle \Psi(0) \vert \cdot \vert \Psi(0) \rangle$. Now further employing a small-angle approximation, $\sin^{2}[\arccos(x)] \simeq 1- x^2$, an approximate expression for the quantum speed limit time $\tau_\mrm{QSL}$ reads
\begin{equation}
\label{eq18}
\tau_\mrm{QSL}^\mrm{lin} \simeq \dfrac{\hbar}{2 E^\mrm{lin}_{\tau}} [1 - \langle \rho^\mrm{in}(\tau) \rangle_{\Psi_0} -\langle \delta\rho(\tau)\rangle_{\Psi_0} ] ,
\end{equation}
and the Hilbert-Schmidt norm of the dynamics becomes
\begin{equation}
\label{eq19}
 E^\mrm{lin}_{\tau} = 1/\tau\,\int_0^\tau dt \, \|H_0\rho^\mrm{in}(t) + \delta\lambda V(t)\rho^\mrm{in}(t) + H_0 \delta\rho(\tau)\|_2\,.
\end{equation}

\paragraph*{Maximal rate of quantum learning.}

Collecting Eqs.~\eqref{eq16}, \eqref{eq18}, and \eqref{eq19} we obtain an expression for the maximal rate of quantum learning, $\Omega$ \eqref{eq06},
\begin{widetext}
\begin{equation}
\label{eq20}
\Omega^\mrm{lin} \equiv 2 E^\mrm{lin}_{\tau}\, \dfrac{\sum_{\alpha} (\Delta\chi^\mrm{in}_{\alpha} + \tr{\delta\rho_\alpha(\tau)} [p^\mrm{in}_\alpha(\tau) - S( \rho^\mrm{in}_\alpha(\tau)) ] + p^\mrm{in}_\alpha(\tau) \tr{\delta\rho_\alpha(\tau) \log  \rho^\mrm{in}_\alpha(\tau)})}{\hbar\, [1 - \langle \rho^\mrm{in}(\tau) \rangle_{\Psi_0} -\langle \delta\rho(\tau)\rangle_{\Psi_0} ] }\,.
\end{equation}
\end{widetext}
Equation~\eqref{eq20} might look mathematically involved, however as we will see in the following, time-dependent perturbation theory allows to analytically solve for $\Omega$ in more complex systems. Observe that only the numerator of $\Omega^\mrm{lin}$ depends on the choice of measurements, whereas the denominator is fully determined by the initial state $\vert \Psi(0)\rangle$. We also note that the eigenvalue energy spectrum governs $\Omega$, which will be further illustrated below by comparing different potentials, namely the harmonic oscillator and the P\"oschl-Teller well.

\paragraph*{Choice of observables.}

To proceed with the analysis we now have to become more specific. Below we will evaluate $\Omega^\mrm{lin}$ for two experimentally relevant systems. Both, the harmonic oscillator and the P\"{o}schl-Teller potential posses symmetric eigenfunction. Thus we choose for illustrative purposes the quantum observable $A$ to measure the symmetry of a quantum state with two outcomes, that we label $e$ for even and $o$ for odd. The corresponding projectors read
\begin{equation}
\label{eq21}
\Pi_e = \sum_n \vert 2n \rangle \langle 2n \vert \quad\text{and}\quad \Pi_o = \sum_n \vert 2n+1 \rangle \langle 2n+1 \vert\,.
\end{equation}
Having chosen the observable, we can now analyze $\Omega$ \eqref{eq06} and $\Omega^\mrm{lin}$ \eqref{eq20} for specific systems.

\subsection{The harmonic oscillator \label{sec:Harm_osc}}

Many phenomena in nature can be described, exactly or approximately by a harmonic potential \cite{Dong2007}. Examples include explaining decoherence in experiments with harmonic optical traps \cite{Wineland2000}, and as testbed for quantum thermodynamics \cite{Deffner2008,Abah2012,Acconcia2015b,Rossnagel2016} and quantum computation \cite{Cory1999}.

The Hamlitonian reads,
\begin{equation}
\label{eq22}
\mathcal{H}(t) = \dfrac{p^2}{2m} + \dfrac{m \omega_0^2}{2}x^2 - m \omega_0^2 \lambda(t) \ x\,,
\end{equation}
with $H_0 = p^2/2m + m \omega_0^2 x^2/2$, and $V(t) = -m \omega_0^2\lambda(t) x$ is the perturbation. For our present purposes $\mathcal{H}(t)$ is particularly elucidating, since its dynamics can be solved analytically \cite{Husimi1953}. Therefore, we can study the range of validity of the linear approximation developed above.

\begin{figure}
\centering
\includegraphics[width=.48\textwidth]{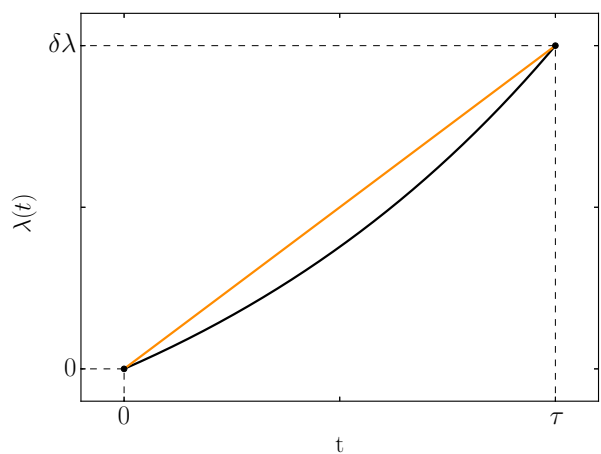}
\caption{(color online) Depiction of the two driving protocols \eqref{eq23}  $\lambda_1(t)$ (black line) and $\lambda_2(t)$ (orange line).}
\label{protocols}
\end{figure}

We will continue for specificity with the two protocols
\begin{equation}
\label{eq23}
\lambda_1(t) = \delta \lambda\,\left(\dfrac{1-e^{t/\tau}}{1-e}\right)\quad\text{and}\quad \lambda_2(t) = \delta \lambda\, \frac{t}{\tau}\ ,
\end{equation}
with $\delta \lambda \ll 1$. These two protocols are depicted in Fig.~\ref{protocols}.
Finally, we assume that the system is initially prepared in its ground state $\vert \Psi(0) \rangle = \vert 0 \rangle $, with corresponding eigenvalue $E_0 = \hbar\omega_0/2$. 

\begin{figure}
\centering
\includegraphics[width=.48\textwidth]{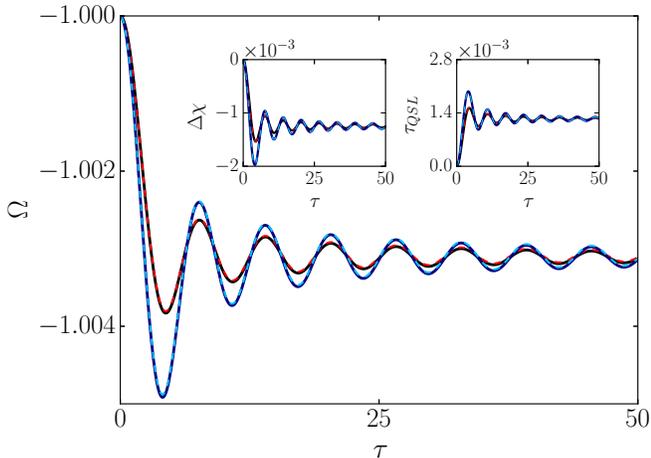}
\caption{(color online) Maximal rate of quantum learning for the time-dependent the quantum harmonic oscillator \eqref{eq22} initially in its
ground state $\vert 0\rangle$. \textit{Insets:} Change of Holevo information (upper left panel), and the quantum speed limit time (upper right panel). Exact results are the black solid line (exponential protocol) and dark blue solid line (linear protocol), and first-order approximations are the red dashed line (exponential protocol) and the light blue dashed line (linear protocol) with parameters $\delta\lambda = 5\times 10^{-2}$ and $\omega_0=1$.}
\label{results_Harm}
\end{figure}

In Appendix \ref{sec:Time_evol_op}, we summarize the analytical solution of the dynamics, which was originally developed in Ref.~\cite{Husimi1953}. Fig.~\ref{results_Harm} summarizes our findings from time-dependent perturbation theory for $\delta\lambda = 5\times 10^{-2}$. We compare linear approximation (dashed lines) with exact solutions  (solid lines) for the exponential protocol $\lambda_1(t)$ as well as for the linear $\lambda_2(t)$ [see Eq.~\eqref{eq23}].

We observe perfect agreement between linear approximation and exact results, which is guaranteed by the aforementioned renormalization procedure \eqref{eq12}. This can be seen, by noting that $\Delta\chi$ is negative for all quench times $\tau$, which is a consequence of the complete positivity of the dynamics. This result is also known as a quantum version of the \textit{data processing lemma} \cite{Nielsen2010}: the information content of a signal cannot be increased by a physical local operation. In other words ``post-processing'' cannot increase information, thus the measurement performed in the end of our process cannot give a $\chi_f  (\tau) > \chi(0)$.

For either parameterization $\Omega$ is finite for infinitely short perturbations, $\tau=0$. For very short quench times we notice an abrupt decay which can be explained by the sudden approximation \cite{messiah1981}: for very rapid changes, the system cannot respond on the same time scale, and the initial state remains unchanged.	For $\tau>0$ we observe strong oscillatory behavior. Even small perturbations are sufficient to induce transitions away from the ground state, which can be attributed to Fermi's golden rule \cite{cohen_vol2_77}. In the limit $\tau\gg 1$, $\Omega$ approaches a constant value, which is determined by the quantum adiabatic approximation \cite{messiah1981}.

The oscillation can be most easily understood from the quantum speed limit time \eqref{eq04}. The sine of the Bures angle, $\sin^{2}[\mathcal{L}]$, between the two states of the harmonic oscillator $\rho(0)$ and $\rho(\tau)$ oscillates approaching a finite value for $\tau \rightarrow \infty$. Consider that the time-evolved density matrix oscillates around the initial state described by $\rho(0)$ due to the small perturbation. For large switching times, the angle tends to a constant value, since in the adiabatic limit no oscillations are present. The period of the oscillations $\Omega$ is inversely proportional to the natural angular frequency of the harmonic oscillator, $T = 2\pi/\omega_0$.  Every minimum value for the maximal rate of learning occurs for specific switching times, in our case in multiples of $2\pi$ given that $\omega_0$ was set to the unit.

Generally, Fig.~\ref{results_Harm} contains experimentally relevant information. For instance, the global minimum of $\Omega$ corresponds to the maximal rate, with which information can be learned about a quantum system evolving in time. We also notice that the two protocols $\lambda_1(t)$ and $\lambda_2(t)$ \eqref{eq23} yield qualitatively similar behavior, but that the linear protocol appears to be more effective, i.e., $|\Omega|$ is larger.

\begin{figure}
\centering
\includegraphics[width=.48\textwidth]{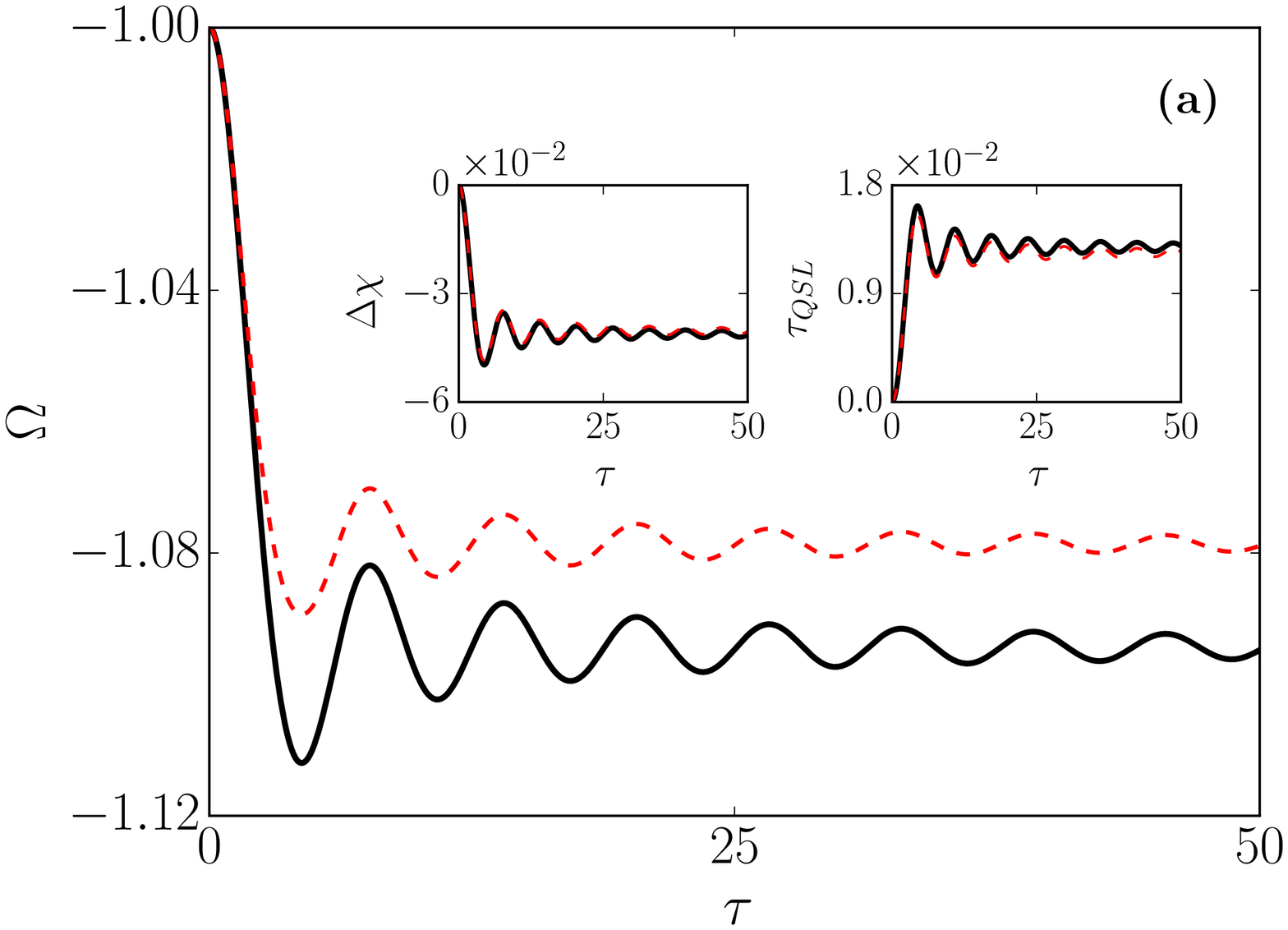}
\includegraphics[width=.48\textwidth]{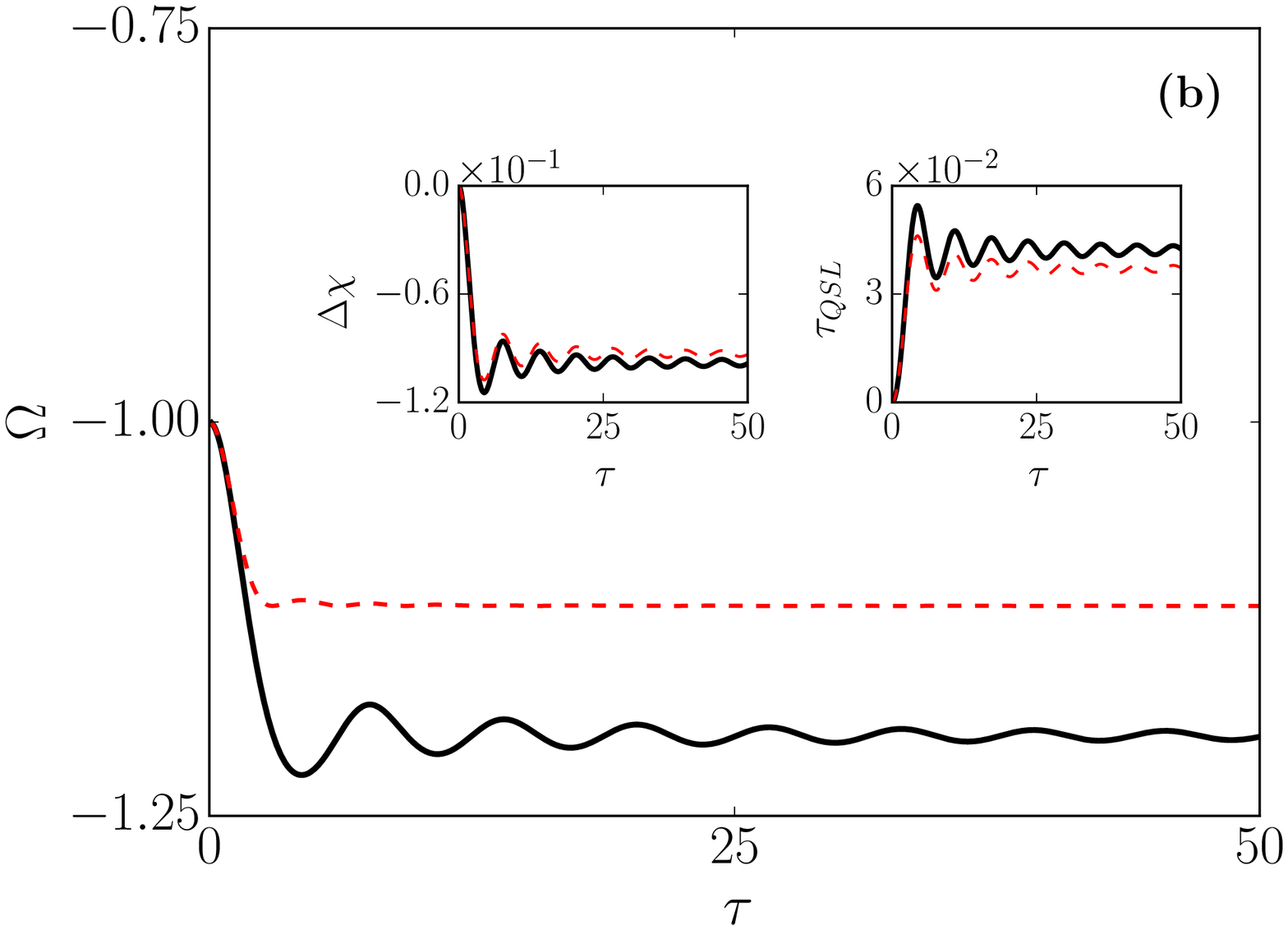}
\caption{(color online) Comparison of the exact (black, solid line) and approximated result (red, dashed line) for the maximal rate of quantum learning for  quantum harmonic oscillator \eqref{eq22} initially in its the ground state $\vert 0 \rangle$. In plot (a) we have $\delta\lambda = 0.3$ and in (b) $\delta\lambda = 0.5$, and as before $\omega_0=1$ and for $\lambda_1(t)$ \eqref{eq23}. \textit{Insets:} Change of Holevo information (upper left panel), $\Delta\chi$, and quantum speed limit $\tau_\mrm{QSL}$ (upper right panel) as a function of the switching time $\tau$.}
\label{maximal_deltas}
\end{figure}

Finally, we also compare our results from perturbation theory with the exact solutions. In Fig.~\ref{maximal_deltas}a we plot $\Omega$ for 
$\delta\lambda = 0.3$ and in Fig.~\ref{maximal_deltas}b for $\delta\lambda = 0.5$. We observe that perturbation theory still gives qualitatively correct behavior, but also that perturbation theory underestimates the magnitude of $\Omega$.

As a main result we have established the relation between the accessible information and dynamic response of the system. These results could prove useful to design optimal strategies to maximize the rate with which information can be retrieved from a quantum system given a particular observable.

\subsection{The P\"{o}schl-Teller potential \label{sec:PT}}

As a second example, we analyze an anharmonic oscillator, which is described by the P\"oschl-Teller potential,
\begin{equation}
\label{eq24}
\mathcal{H}(t) = \dfrac{p^{2}}{2m} - \frac{1}{2}\,\nu(\nu+1)\, \mathrm{sech}^2(x) - \eta\lambda(t)\, x \,,
\end{equation}
where $\eta$ is a constant. 

The P\"{o}schl-Teller potential was originally introduced to study vibrational excitations in polyatomic molecules \cite{Poschl1933}. Since then it has been applied to describe a wide variety of processes ranging from neutron scattering \cite{Carlson2010} to many-body systems \cite{Pacheco2001}, and in the description of symmetries of spin-orbit coupling for quantum relativistic systems \cite{Jia2008,Jia2009,Jia2010}. On the experimental side, Eq.~\eqref{eq24} has proven useful in quantum optics \cite{Lekner2007} and to describe different refraction indices according to the parameters of the setup \cite{Tomak2006}. Moreover, Eq.~\eqref{eq24} has been used in the laboratory to describe quantum dots in semiconductors nanoelectronics and in the modeling of optoelecetronic devices \cite{Hayrapetyan2013,Hayrapetyan2014}.  

In contrast to the time-dependent harmonic oscillator \eqref{eq22}, the dynamics of P\"oschl-Teller potential \eqref{eq24} is not analytically known . Therefore, we have to rely on our results from time-dependent perturbation theory \eqref{eq20}. For the numerical analysis we chose $\nu = 20$ and set as the initial state $\vert \Psi(0)\rangle = \vert 10 \rangle$. Moreover, $\Omega^\mrm{lin}$ was computed again for the parity as observable \eqref{eq21} and the driving protocols in Eq.~\eqref{eq23} with $\delta\lambda = 0.1$.
\begin{figure}
\centering
\includegraphics[width=.48\textwidth]{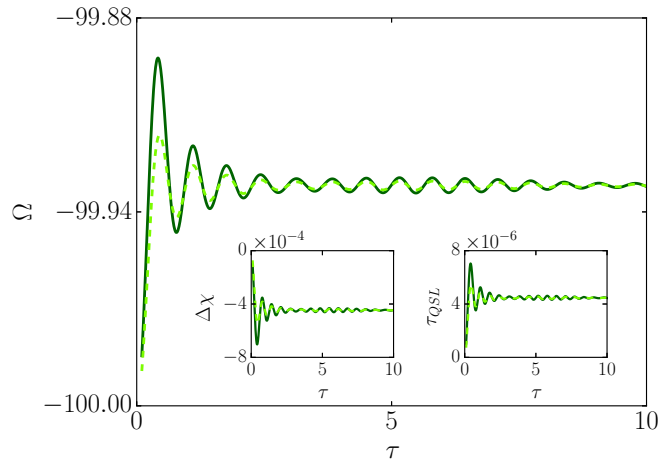}
\caption{(color online) $\Omega^\mrm{lin}$ \eqref{eq20} for the time-dependent P\"{o}schl-Teller potential \eqref{eq24} with $\vert \Psi(0)\rangle = \vert 10\rangle$, $\nu=20$, and $\eta=1$. \textit{Insets:} Change of Holevo information (lower left panel), and the quantum speed limit time (lower right panel). Results are for $\lambda_1(t)$ \eqref{eq23} (light green, solid line) and for $\lambda_2(t)$ \eqref{eq23} (dark green, solid line) with $\delta\lambda = 0.1$.}
\label{result_pt}
\end{figure}	

Our results are summarized in Fig.~\ref{result_pt}. Similar to the harmonic oscillator, the short time behavior is fully characterized by the sudden approximation, and by the adiabatic approximation for long quench times $\tau$. However, we also observe that the oscillations are smaller, which suggests that the P\"oschl-Teller potential is less susceptible to the specific driving protocol. Overall, however, our findings for the harmonic oscillator \eqref{eq22} in Fig.~\ref{results_Harm} and for the  P\"oschl-Teller potential \eqref{eq24} in Fig.~\ref{result_pt} are remarkably similar.

\paragraph*{Comparison of the systems.}

\begin{figure}
\centering
\includegraphics[width=.48\textwidth]{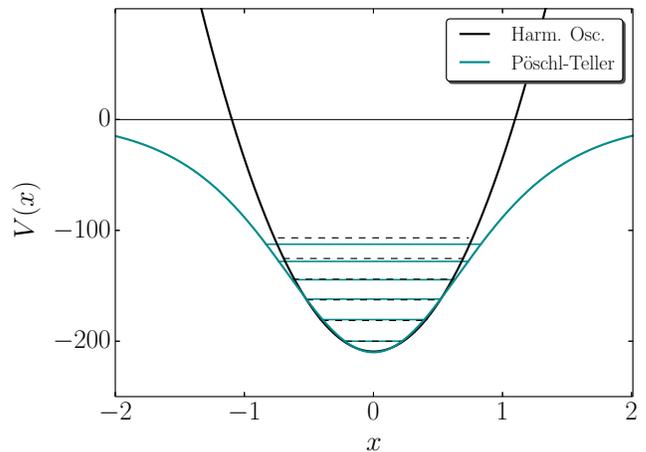}
\caption{(color online) Harmonic approximation of the P\"{o}schl-Teller potential. The black dashed lines are the first five energy eigenenergies of the harmonic oscillator (black line) with $\omega_0 = 18.65$ shifted by $E_c = - 209.325$; blue solid lines represent the first five eigenenergies for the P\"oschl-Teller potential \eqref{eq22} with $\nu = 20$.}
\label{potentials}
\end{figure}

For the chosen parameter, $\nu=20$, the low-lying eigenenergies are well approximated by an harmonic oscillator of $\omega_0 = 18.65$ and where the ground state energy is shifted by $E_c = -209.325$, see Fig.~\ref{potentials}. Thus we would expect similar behavior for the maximal rate quantum learning $\Omega^\mrm{lin}$ \eqref{eq20}.

\begin{figure}
\centering
\includegraphics[width=.48\textwidth]{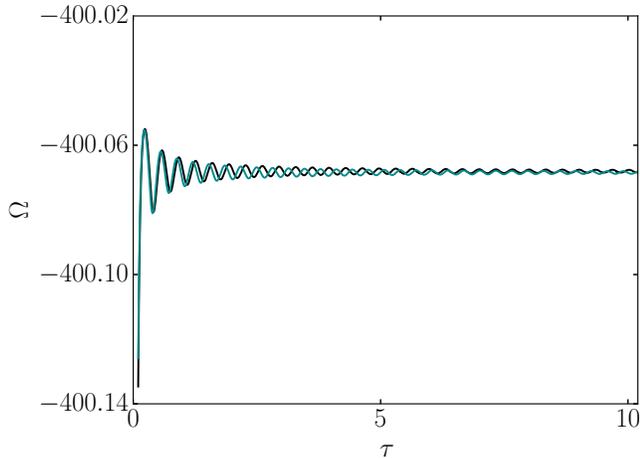}
\caption{(color online) Comparison of $\Omega^\mrm{lin}$ \eqref{eq20} for the P\"{o}schl-Teller potential (blue line) and the harmonic oscillator (black line).}
\label{maximal_comp}
\end{figure}

Above we have seen that these two potentials appear not be equally susceptible to perturbations. This, however, is not the case. The different behavior of $\Omega^\mrm{lin}$ and in particular the size of the fluctuations are governed by the choice of the initial state. If both harmonic oscillator and P\"oschl-Teller potential are initialized in the corresponding ground states with the same eigenenergy the resulting $\Omega^\mrm{lin}$ is almost identical. Correcting for the relative magnitude of the perturbation, we choose $\delta\lambda_{HO} = 2.87 \times 10^{-4}$ for the harmonic and $\delta\lambda_{PT} = 10^{-1}$ for the P\"oschl-Teller potential. The numerical results for the exponential protocol $\lambda_1(t)$ \eqref{eq23} can be found in Fig.~\ref{maximal_comp}.

We observe that for short quench times and for long quench times the agreement is almost perfect. For intermediate quench time, however, the P\"oschl-Teller potential and its harmonic approximation are ``out of sync''. This difference evidences of the nonlinear (nonharmonic) properties of the P\"oschl-Teller potential \eqref{eq24}.

\section{Concluding remarks \label{sec:conclusions}}

In this analysis we have obtained three main results: (i) we have revisited the Bremermann-Bekenstein bound and generalized the maximal rate of information transmission to account for the effect of measurements; (ii) our new bound can be interpreted as a maximal rate of quantum learning, which presents an important application of the quantum speed limit time; and (iii) we have computed approximate expressions from time-dependent perturbation theory, which we are able to show to be exact for small enough perturbations.

Our results were illustrated for two experimentally relevant systems, namely the driven harmonic oscillator and the P\"oschl-Teller potential. In particular, our comparison of the maximal rate of quantum learning for the P\"oschl-Teller potential and its harmonic approximation suggests a wide range of applicability also for more complex systems. However, for the present purposes and for pedagogical reasons we chose simple systems, analytical driving protocols, and tractable observables.

\paragraph*{Next steps.} A more realistic approach would take into account the fact that electronic devices processes information manipulating electric currents. In addition, it would be interesting to investigate how our present results  have to modified, if we account for the relativistic nature of electrons.

\begin{acknowledgments}
It is a pleasure to thank V. L. Quito and L. M. M. Dur\~{a}o for inspiring and fruitful discussions. T.A. acknowledges support from ‘Gleb Wataghin’ Physics Institute (Brazil) and CAPES (Brazil), Grant No.  1504869, and the hospitality of the Center for Nonlinear Studies at the Los Alamos National Laboratory, where this project was conceived.  S.D. acknowledges financial support by the U.S. National Science Foundation under Grant No. CHE-1648973, and the hospitality of the Universidade Estadual de Campinas, where this project was concluded.
\end{acknowledgments}

\appendix

\section{Time evolution operator \label{sec:Time_evol_op}}

In this appendix we outline the analytical solution of the dynamics induced by the time-dependent harmonic oscillator \eqref{eq22}.
To begin, we write the Hamiltonian $\mathcal{H}(t)$ subject to an external perturbation $V(t)$ in interaction picture
\begin{equation}
\label{a01}
\begin{split}
\mathcal{H}_I(t)& = e^{i H_0 t/\hbar} (H_0 + V(t)) e^{-i H_0 t/\hbar}\\
& = H_0+e^{i H_0 t/\hbar} \ V(t)\  e^{-i H_0 t/\hbar} \ ,
\end{split}
\end{equation} 
and  consequently we write the time-dependent wave function as $\Psi_I(t) = U_I(t,t_0)\Psi_I(t_0)$.  

\paragraph*{Magnus expansion.}The time-evolution operator can be written as a solution of 
\begin{equation}
\label{a02}
i \hbar \dfrac{dU(t,t_0)}{dt} = \mathcal{H}(t) U(t,t_0) \,.
\end{equation}
Such linear differential equations of the form $dY(t)/dt = A(t)Y(t)$, with the inital value $Y(t_0)=Y_0$ and $A(t)$ being a $n\times n$ matrix with time-varying coefficients, can be solved using the so-called Magnus expansion \cite{Magnus1954}. Generally, we can thus write
\begin{equation}
\label{a03}
\dfrac{d e^{\Omega}}{dt}  e^{-\Omega} = A(t)\ ,
\end{equation}
where $\exp(\Omega(t,t_0))$ is expressed as an infinite series: $\exp(\Omega(t,t_0)) = \sum_k \Omega_k (t,t_0)$. Magnus noticed that $A(t)$ can be found using a Poincar\'{e}-Hausdorff matrix identity, and also that the problem can be solved relating the time derivative of $\Omega(t,t_0)$ with other special functions. In Ref.~\cite{Pechukas1966} this expansion is used, and we identify $Y(t)$ as the time-evolution operator $U(t,t_0)$ and $A(t)$ is the time-dependent Hamiltonian $\mathcal{H}(t)$. After some lines of algebra one can show \cite{Pechukas1966}	
\begin{widetext}
\begin{equation}
\label{a04}
\begin{split}
A^{'}_{n+1} = \left\lbrace - \dfrac{C_{A_{n}}}{2} + \dfrac{B_1}{2} (C_{A_1}C_{A_{n-1}} + C_{A_2}C_{A_{n-2}} + \cdots) \right\rbrace H(t)\ ,
\end{split}
\end{equation}
where $A^{'}_{n}$ is the derivative of the $n$th element of the expansion of $A(t)$, $C_{A}B = [A,B]$, and $B_{i}$'s are the Bernoulli numbers. It is interesting to note that this formulation respects the time ordering of the operators. Finally, each term $A_n$ can be obtained by integrating over time in Eq.~\eqref{a04}. 

Now, recognizing $\exp[A(t,t_0)] = U(t,t_0)$, we have for the time-evolution operator in interaction picture
\begin{equation}
\label{a05}
U_I(t,t_0) = \exp\left( \sum_{n} A_n \right) = \exp\left(\int_{t_{0}}^t \dfrac{\mathcal{H}_I(t_1)}{i\hbar} dt_1 - \dfrac{1}{2} \int_{t_{0}}^t dt_{2} \int_{t_{0}}^{t_2} \left[\dfrac{\mathcal{H}_I(t_1)}{i\hbar}, \dfrac{\mathcal{H}_I(t_2)}{i\hbar}\right] dt_1  \ + \cdots \right)\,..
\end{equation}
It is the easy to see that for the harmonic oscillator \eqref{eq22} we also have
\begin{equation}
\label{a06}
[\mathcal{H}(t_1), \mathcal{H}(t_2)] = 2 i f(t_1) \ f(t_2) \ \sin(\omega_0(t_1-t_2))\ ,
\end{equation}
where $f(t) = - \sqrt{m \omega_0^3 \hbar /2} \ \lambda(t)$.

Thus, we obtain the exact expression for $U(t,t_0)$ in the Schr\"{o}dinger picture
\begin{equation}
\label{a07}
U(t,t_0) = \left(\exp[i\beta]\exp[i (\alpha a^{\dagger} + \alpha^{*}a - \gamma)]\right) \exp[-iH_0t/\hbar]\ ,
\end{equation}	
where $a^{\dagger}$ and $a$ are the creation and annihilation operators respectively, and the coefficients are
\begin{eqnarray}
\alpha &=&  \sqrt{\dfrac{m\omega_0^{3}}{2\hbar}} \int_{t_0}^{t} dt^{'} \ e^{i\omega_0 t^{'}} \lambda(t^{'}) \ ,\\
\beta &=& - \dfrac{m\omega_0^{3}}{2\hbar} \int_{t_{0}}^t dt_{2} \int_{t_{0}}^{t_2} dt_1 \ \lambda(t_1) \ \lambda(t_2) \sin(\omega_0(t_1-t_2)) \ ,\\
\gamma &=& \dfrac{m\omega_0^{2}}{2\hbar} \int_{t_0}^{t} dt^{'} \ \lambda^2(t^{'}) \ .
\end{eqnarray}
Note that as $\lambda(t) \rightarrow 0$ all terms $\alpha$, $\beta$ and $\gamma$ vanish, and we recover the time evolution operator for the non-perturbed harmonic oscillator, $U(t_0,t) \rightarrow \mathrm{exp}[-iH_0t/\hbar]$.

\paragraph*{Method of generating functions.} An alternative solution was proposed by Husimi \cite{Husimi1953}. In his approach one writes the time-dependent solution as $\Psi(x,t) = \sum_n c_n(t) e^{-i H_0 t/\hbar} \Psi_n(x)$, and we have
\begin{equation}
\label{a07b}
i \hbar \dfrac{d c_m}{dt} = \sum^{\infty}_{n=0} c_n f(t) e^{i (m-n)\omega_0 t} x_{mn} \quad  \text{and}\quad\ i \hbar \dfrac{d c(t)}{dt} = \mathcal{H}(t)\ c(t) \ ,
\end{equation}
where $x_{mn} = \int \Psi_m x \Psi_n$, and $\mathcal{H}_{mn}(t) = - f(t) e^{i (m-n)\omega_0 t} x_{mn}$ are the elements of the matrix $\mathcal{H}(t)$.

The latter can be solved by the method of iteration, and we find a formal solution as an infinite series,
\begin{equation}
\label{a08}
c(t) = c(0) \left(1 + \dfrac{1}{i \hbar} \int_{0}^{t} \mathcal{H}(t_1) \ dt_1 + \left(\dfrac{1}{i \hbar}\right)^2 \int_{0}^{t} \mathcal{H}(t_1) \ dt_1 \int_{0}^{t1} \mathcal{H}(t_2) \ dt_2 + \cdots \right)  = S(t,0) c(0)\ .
\end{equation}
The $n$th term of $S(t,0)$ reads
\begin{equation}
\label{a09}
S_n(t,0) = \dfrac{(-1)^n}{n!} \int_0^t \cdots \int_0^t dt_1 \ dt_2  \cdots dt_n \ \mathcal{T}_>(\mathcal{H}(t_1)\mathcal{H}(t_2) \cdots \mathcal{H}(t_n) )
\end{equation}
where $\mathcal{T}_>$ is the time-ordering operator. 

Using Wick's theorem to write the time-ordered products in Eq.~\eqref{a09} in terms of normal ordered products, simplifies the problem for operators that annihilates the vacuum. The time-ordered operator in Eq.~\eqref{a09} can be written as contractions of field operators, which are commutators \cite{Ticciati1999,Molinari2016},
\begin{equation}
\label{a10}
\mathcal{T}_>(A(t_1)B(t_2)) = [A^{+}(t_1),B^{-}(t_2)] \ + :A(t_1)B(t_2): 
\end{equation}
where $:AB:$ is the normal order for the field operators. We have used $t_1>t_2$ in the last equation.

Now, writing the Hamiltonian in interaction picture, $\mc{H}(t)$ can be expressed as a function of $V(t)$. The time-ordering in Eq.~\eqref{a10} will be applied over the perturbation which is a proportional to the creation and annihilation operators, $a(t_i)$ and $a^{\dagger}(t_j)$. In conclusion, we have that Husimi's approach is fully equivalent to Ref.~\cite{Pechukas1966}.

Thus, we can also write \cite{Husimi1953}
\begin{equation}
\label{a11}
U_{mn}(t,t_0) = \dfrac{e^{-i(t-t_0)/2 + i \sigma/2}}{\sqrt{2^{n+m} m! n!}} e^{-W/2} \ \mathcal{C}(m,n\vert W) (\xi + i \dot{\xi})^{m} (\eta - i\eta^{'})^{n} 
\end{equation}
where the coefficients are given by
\begin{eqnarray}
f(t) &=& -m\omega_0^2 x \lambda(t) \ , \\
\sigma &=& \int_0^t dt^{'} \int_{t_0}^{t^{'}} dt^{''} \sin(t^{'}-t^{''}) f(t^{'}) f(t^{''}) \ ,\\
W &=& \dfrac{1}{2}  \int_{t_{0}}^t  \int_{t_0}^{t} \cos(t^{'}-t^{''}) f(t^{'}) f(t^{''}) dt^{'} dt^{''} \ ,\\
\mathcal{C}(m,n\vert W) &=& \sum_{l=0}^{\min m,n} \dfrac{m! \ n!}{l! (m-l)! (n-l)!} (-W)^{-l}\ , \\
\xi + i \dot{\xi} &=&  -i e^{-i t} \int_0^{t} du \ e^{i u} f(u) \ , \\
\eta - i\eta^{'} &=& -i \int_{0}^{t} du \ e^{iu} f(u) \ .
\end{eqnarray}
\end{widetext}

\bibliography{paper_Ref}
\end{document}